\documentclass[aps]{revtex4}
\usepackage{amsmath}
\usepackage{epsfig, graphicx}
\newcommand{\nuc}[2]{\ensuremath{^{#2}}\textrm{#1}}

\begin{document}

\title{Determining Reactor Flux from Xenon-136 and Cesium-135 in Spent Fuel}

\author{A.C. Hayes and Gerard Jungman}
\affiliation{Los Alamos National Laboratory, Los Alamos, NM  87545}
\email{anna_hayes@lanl.gov}
\hfill \fbox{\parbox[t]{1.12in}{LA-UR-12-21011}}\hspace*{0.35in}

\date{\today}

\begin{abstract}
The ability to infer the reactor flux from spent fuel or seized fissile material would enhance the tools of
nuclear forensics and nuclear nonproliferation significantly. We show that reactor flux can be inferred from
the ratios of xenon-136 to xenon-134 and cesium-135 to cesium-137. If the average flux of a reactor is known, 
the flux inferred from measurements of spent fuel could help determine whether that spent fuel was loaded as a blanket or close to the mid-plane of the reactor. 
The cesium ratio also provides information on
reactor shutdowns during the irradiation of fuel, which could prove valuable for identifying the reactor in
question through comparisons with satellite reactor heat monitoring data. 
We derive analytic expressions for these correlations and compare them to experimental data and
to detailed reactor burn simulations. 
The enrichment of the original uranium fuel affects the correlations
by up to 3 percent, but only at high flux.
\end{abstract}

\maketitle

\section{Introduction}
Verification of reactor operations is an important component of nuclear nonproliferation and safeguards.
Of particular importance is the neutron flux to which fuel was exposed during reactor operation, which
might not necessarily be the same as the average flux quoted for a reactor. For example, if the spent
fuel in question was loaded in the reactor as a blanket, it would generally see a lower flux than
would the fuel loaded in the mid-plane. Thus, it is important to verify the flux from measurements
of the spent fuel itself.
Ratios of fission products in spent reactor fuel or released as fission gases during fuel reprocessing are used to
deduce information about the irradiation history of the fuel. In the case of seized undetonated nuclear material,
fission product ratios can enable inferences about the reactor used to generate the fuel, particularly when compared with reactor monitoring information.
In this paper, we examine the correlation between the \nuc{Xe}{136}/\nuc{Xe}{134}  and  \nuc{Cs}{135}/\nuc{Cs}{137} fission
 product ratios and the thermal neutron flux used to irradiate the fuel. We also examine how the reactor power
 history, particularly reactor shutdowns, might
also be extracted from these ratios.
Correlations between a number of fission products and the reactor operations were observed experimentally by Maeck
{\it et al.} \cite{Maeck}. In the present work we identify the underlying physics governing the dependence
of the xenon and cesium ratios on reactor operation, we derive functional forms for these ratios, and we
compare these with detailed reactor simulations and with the experimental data. 

The sensitivity of the two fission product ratios to the reactor thermal neutron flux arises because of the
competition between the 9.14 hour decay of xenon-135 to cesium-135  and the thermal neutron capture on xenon-135
to xenon-136, with an abnormally large cross section of $2.6\times 10^6$ barns, Figure 1.
For  low flux, the majority of xenon-135  produced in fission \(\beta\)-decays to cesium-135, 
while for high flux most of the xenon-135 is transmuted to xenon-136.
In contrast, reactor production of xenon-134 and cesium-137 is simply dependent on the number of
fissions, because both of these nuclei are produced almost entirely as
direct fission products or by the beta decay of other fission products. 

The accuracy with which the cesium and xenon ratios can be measured depends on the size of the sample. 
Using  resonance ionization mass spectrometry, Pibida {\it et al.} \cite{Pibida} have shown that the 
\nuc{Cs}{135}/\nuc{Cs}{137} ratio can be determined reliably to 1\% accuracy in very low-level samples
containing large natural cesium backgrounds. 
This cesium ratio has been used \cite{Taylor} as a forensic tool for  identifying the source of  radioactive
contamination in soil samples, though there is some evidence \cite{synder} to suggest that isotope fractionation
may  affect the cesium ratio in reactor effluences, depending on the location of the measured sample.  
Resonance ionization mass spectrometry has been shown \cite{gilmour} to be capable of determining
ratios of naturally occurring xenon isotopes to better than 0.5\%  in samples of $10^{-12}$ $cm^3$ STP of xenon. 
For large samples, mass spectrometry of both xenon and cesium ratios can achieve an accuracy better than $1:10^4$.

\begin{figure}
\includegraphics[width=12.5cm]{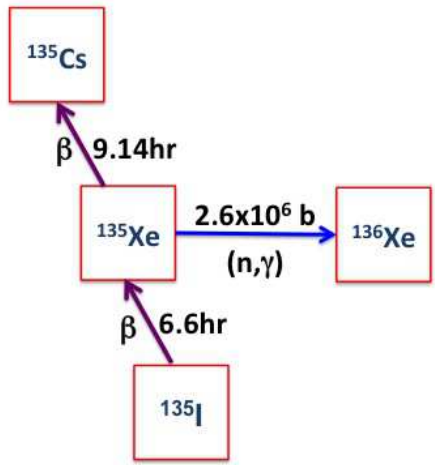}
\caption{The competition between the decay of xenon-135 to cesium-135
and the thermal neutron capture on xenon-135 to xenon-136 causes the relative concentrations of xenon-136 and
cesium-135 in reactor fuel to be sensitive indicators of the thermal neutron flux.} 
\label{135-Xe}
\end{figure}

The concentration of xenon-135 under different reactor conditions 
is well studied \cite{book} because of the role it plays as a serious reactor poison.
If the reactor is operating in a steady state with a constant neutron flux, the concentration of xenon-135
reaches an equilibrium value that depends on the concentration of iodine-135  and on the rate of capture to
xenon-136.
Changes in the thermal flux cause short-term fluctuations in the level of xenon-135, and after about 40 to 50 hours
of steady flux, the xenon-135 level settles to a new equilibrium value that reflects the new flux.
Such changes in flux, as well as shutdowns and restarts of the reactor, also affect the \nuc{Xe}{136}/\nuc{Xe}{134}
and \nuc{Cs}{135}/\nuc{Cs}{137} ratios.
In this paper we quantify these effects and show how measurements of these two ratios could provide
important information on reactor flux and power history.

\section{The Dependence of \nuc{Xe}{136} on the thermal flux}

Xenon-136 is produced by three main mechanisms in a reactor: 
as a direct fission product,  from the $\beta$-decay of the fission fragment iodine-136,  
and from neutron capture on xenon-135. 
The lifetimes of the nuclei in the beta-decay chain leading to xenon-136 are short, and the first two mechanisms
can be co-added and described in terms  of the so-called cumulative fission yield for xenon-136. The cumulative
fission yields for all fission products have been evaluated by England and Rider \cite{england} and
Wahl \cite{Wahl}; the cumulative fission yield for xenon-136 is $\sim7$ percent (\%) per fission for the
fissioning isotopes of uranium and plutonium. 

Under steady state conditions, the growth rate of xenon-136 in the reactor is, 
\begin{eqnarray}
\dot{N}_{136Xe}& =& \overline{f}_{136Xe} \Gamma_F+ N_{135Xe}^{equil}\phi_T\sigma_a\\
&=& 
\overline{f}_{136Xe}\Gamma_F \left[ 1 
+ \left( \frac{\overline{f}_{135Xe}}{\overline{f}_{136Xe}}\right)
\frac{\phi_T\sigma_a}{\lambda_{135I}+\phi_T\sigma_a}\right],
\label{136-1}
\end{eqnarray} 
where $\Gamma_F$ is the fission rate, $\overline{f}_A$ is the burn-weighted cumulative fission
yield of nucleus $A$
\footnotetext[1]{The cumulative yields $f_A$ are actinide dependent. Throughout this paper we use the
notation $\overline{f}_A$ to mean the cumulative yield for fragment $A$ when weighted by the linear
combination of burning actinides. Since the linear combination of actinides contributing to the burn
changes with time, in general, the cumulative yields $\overline{f}_A$ are also time dependent.}, 
$\phi_T$ is the thermal neutron flux, $\sigma_a$ the thermal capture cross section on xenon-135,
and $N_{135Xe}^{equil}$ is the equilibrium value of xenon-135.

The dependence of the xenon-136 production rate on the
magnitude of the thermal neutron flux falls between two limits for all thermal reactors,
\begin{eqnarray}
 \phi_T\sigma_a\gg\lambda_{135}:\; \dot{N}_{136}& = &\overline{f}_{136} \Gamma_F(1+\frac{\overline{f}_{135}}{\overline{f}_{136}})\\\nonumber
 \phi_T\sigma_a\ll\lambda_{135}:\; \dot{N}_{136}& = &\overline{f}_{136} \Gamma_F 
\label{limits}
\end{eqnarray}

\begin{figure}
\vspace{0.8 cm}
\includegraphics[width = 8.5cm]{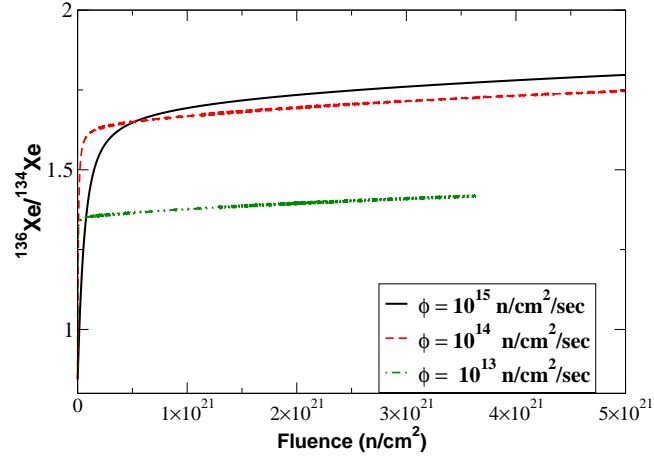}
\caption{The \nuc{Xe}{136}/\nuc{Xe}{134} ratio equilibrates to a value that is determined by the thermal
neutron flux, Equation (\ref{136-134}). Thus, this ratio can be used to deduce the flux. The residual
slope of the plateaus in the curves is caused by the slow evolution of the fuel composition, which is
not explicitly taken account in Equation (\ref{136-134}), except through the implied evolution of the cumulative yields $\overline{f}_A$; see the discussion for Figure \ref{xe-enrich}.}
\label{136-134-a}
\end{figure}

The production rate of xenon-134 only depends on the fission rate 
\footnotetext[2]{If the fuel is irradiate for a long time $\sim 1$ year, 
production of all most isotopes by neutron capture becomes non-negligible. In the present work we
are assuming that the burn times of interest are low enough to ignore neutron capture, except in the
case of capture on xenon-135.} 
$\dot{N}_{134}=\overline{f}_{134}\Gamma_F$.
Thus,
\begin{equation}
\frac{N_{136}}{N_{134}} = \frac{\overline{f}_{136Xe}}{\overline{f}_{134Xe}} 
\left(1
+ \left( \frac{\overline{f}_{135Xe}}{\overline{f}_{136Xe}}\right)
\frac{\phi_T\sigma_a}{\lambda_{135}+\phi_T\sigma_a}\right)\;.
\label{136-134}
\end{equation}
From Equation (\ref{136-134}), the \nuc{Xe}{136}/\nuc{Xe}{134} ratio equilibrates on a (flux dependent) time
scale of weeks. Figure \ref{136-134-a} displays the situation for different values of the thermal flux.

\subsection{Power Fluctuations and the Average Flux from \nuc{Xe}{136}/\nuc{Xe}{134}}

\begin{figure}
\vspace{0.8 cm}
\includegraphics[width = 8.5cm]{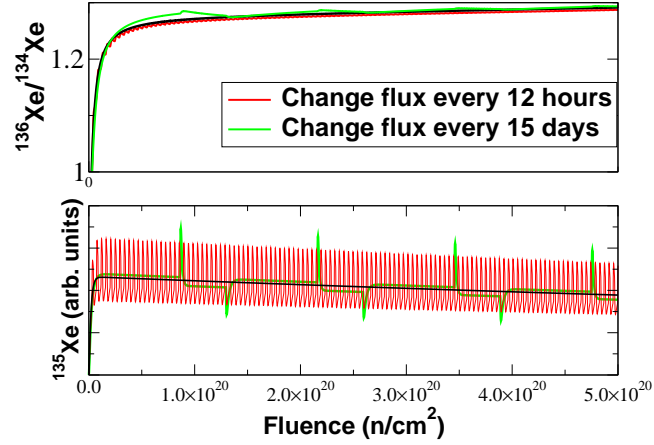}
\caption{Changes in the concentration of xenon-135 and in the \nuc{Xe}{136}/\nuc{Xe}{134} ratio with fluctuations in
the thermal power. The red line corresponds to a change in the thermal flux by a factor of two every 12 hours,
and the green line to change in the flux by a factor of two every 15 days. In both cases, the flux was varied
between $5\times10^{12}\; n/cm^2/sec$ and $10^{13}\; n/cm^2/sec$. The black line corresponds to a constant
flux of $7.5\times10^{12}\; n/cm^2/sec$, which compares well with the results obtained with the fluctuating flux.
Thus, with variations in the flux, the \nuc{Xe}{136}/\nuc{Xe}{134} ratio determines the {\it average} flux,
which for most applications is the physical flux of interest.}
\label{vary}
\end{figure}

If the reactor power is varied through variations in the thermal flux, the concentration of xenon-135
adjusts and reaches a new level.
The value of the  \nuc{Xe}{136}/\nuc{Xe}{134} ratio also reflects these variations.
For this reason, previous studies \cite {Nakhleh,Charlton,Feary,Burr}  of xenon gases released during fuel
reprocessing, for the purposes of extracting reactor operation  information, omitted xenon-136 in their
analyses.
The concern was that, without knowledge of the power history, xenon-136 would be difficult to interpret.
However, the  \nuc{Xe}{136}/\nuc{Xe}{134} ratio is  a direct 
measure of the {\it  average flux} over the burn of the fuel. 
To show this we ran a series of simulations in which we varied the flux over different time periods.
Figure \ref{vary} shows the effect on the concentration of xenon-135 of varying the power (1) on a daily basis
and (2) every 15 days. In these calculations we kept the thermal neutron flux
fixed at $\phi_T = 10^{13}\; n/cm^2/sec$ for the first 12 hours each day (15 days of each month) and dropped it to
$\phi_T = 0.5\times10^{13}\; n/cm^2/sec$ for the second 12 hours (15 days).
This caused the xenon-135 concentration to vary up and down, as displayed by the red (green) oscillating lines
in the lower panel of Figure \ref{vary}. 
The black line shows the effect of keeping the flux fixed at the average of these fluctuations,
$\phi_T =7.5\times 10^{12}\; n/cm^2/sec$. The upper panel 
of Figure \ref{vary} shows the resulting \nuc{Xe}{136}/\nuc{Xe}{134} ratio, where it is clear that, in the
presence of power variations, the  \nuc{Xe}{136}/\nuc{Xe}{134} ratio determines the  average thermal flux.
This average flux is the physically meaningful quantity in determining flux-dependent
physical properties of the spent fuel. At very high flux values, the nonlinearity in
Equation (\ref{136-134}) begins to skew the the relation to the average flux, but this
systematic effect does not become important until the flux is significantly higher
than \(10^{14}\; n/cm^2/sec\).

\section{The relationship between thermal flux and the \nuc{Cs}{135}/\nuc{Cs}{137} ratio}
In this section we derive an analytic expression relating the concentration of  cesium-135 to the thermal
neutron flux.
As seen in Figure 1, cesium-135 is produced through the decay chain
\(\nuc{I}{135} \rightarrow \nuc{Xe}{135} \rightarrow \nuc{Cs}{135}\), which is complicated by the
transmutation of xenon-135 via neutron capture. 
The number of cesium-135 atoms produced in a time \(t\) is given by

\begin{equation}
N_{135Cs}(t) = \lambda_{135Xe}\int_0^t dt' 
\left[\frac{\frac{\overline{f}_{135I}}{\overline{f}_{137Cs}}\dot{N}_{137Cs}-\dot{N}_{135I}-\dot{N}_{135Xe}}{\lambda_{135Xe}+\phi_T\sigma_a}\right].
\label{tintegral}
\end{equation}
Here  \(\dot{N}_{A}\) is the growth rate of nuclide A.

To solve for the \nuc{Cs}{135}/\nuc{Cs}{137} ratio exactly requires numerical reactor burn simulations, which we
present below. However, an analytic expression for this ratio can be obtained by assuming that, while the
reactor is operating,  the concentration of iodine-135 and xenon-135 are at their equilibrium values. 
We also assume that the measurements of cesium-135 are made after all of the iodine-135 and xenon-135 in the
fuel has decayed.
Thus, if the reactor is run with constant flux we can write
\begin{eqnarray}
N_{135Cs} &= &\frac{\lambda_{135Xe}}{\lambda_{135Xe}+\phi_T\sigma_a}\frac{\overline{f}_{135I}}{\overline{f}_{137Cs}}N_{137Cs} +  \frac{\phi_T\sigma_a}{\lambda_{135Xe}+\phi_T\sigma_a}\left(N_{135I}^{equil}+N_{135Xe}^{equil}\right)\\
& = & N_{137Cs} \frac{\overline{f}_{135I}}{\overline{f}_{137Cs}}\left [\frac{\lambda_{135Xe}}{\lambda_{135Xe}+\phi_T\sigma_a} + \frac{\phi_T\sigma_a}{\lambda_{135Xe}+\phi_T\sigma_a}\left(\frac{1}{\lambda_{135I}T_{irrad}}\right)\left(1+\frac{\lambda_{135I}}{\lambda_{135Xe}+\phi_T\sigma_a}\right)\right].
\label{once}
\end{eqnarray}

Here $\lambda_{135I}$ is the decay constant for iodine-135, and 
$T_{irrad}$ is the total irradiation time in the reactor, which we introduced by making the substitution,
$N_{137Cs}=\overline{f}_{137Cs}\Gamma_F\;T_{irrad}$. The form of the second term, which gives rise to the term
inversely proportional to the total irradiation time, arises from the proper treatment of startup transients
in integrating Equation (\ref{tintegral}).

\subsection{Effect of Reactor Shutdowns on the \nuc{Cs}{135}/\nuc{Cs}{137} Ratio}

We next consider the scenario in which the reactor is shutdown once or several times and restarted.
Because xenon-135 is a serious reactor poison, shutdown times  normally have to be long enough for all of the
xenon-135 (and iodine-135) to decay. Thus, each time the reactor is shutdown the cesium-135 concentration is
increased by the \nuc{I}{135} $\rightarrow \nuc{Xe}{135} \rightarrow$ \nuc{Cs}{135} decay chain.
Assuming that the flux is constant during operating periods, the effect of shutdowns is simply to  add  terms
identical to the second term in Equation (\ref{once}), one for each time the reactor is shutdown and restarted.
If $P$ represents the number of times the reactor is run and shutoff before the fuel is removed, with
$P_{minimum}=1$, the \nuc{Cs}{135}/\nuc{Cs}{137} ratio is given by,
\begin{equation}
N_{135Cs}/N_{137Cs} = \frac{\overline{f}_{135I}}{\overline{f}_{137Cs}}\left [\frac{\lambda_{135Xe}}{\lambda_{135Xe}+\phi_T\sigma_a} + \frac{\phi_T\sigma_a
}{\lambda_{135Xe}+\phi_T\sigma_a}\left(\frac{P}{\lambda_{135I}T^{total}_{irrad}}\right)\left(1+\frac{\lambda_{135I}}{\lambda_{135Xe}+\phi_T\sigma_a}\right)\right]\;.
\label{cesium-ans}
\end{equation} 
Here $T^{total}_{irrad}$ is the sum of all the irradiation times to which the fuel was exposed.
Comparing the magnitude of the main contribution to that of the correction to the \nuc{Cs}{135}/\nuc{Cs}{137} ratio, 
it can be seen that the correction becomes significant for high flux.
Thus, a measurement of the  \nuc{Cs}{135}/\nuc{Cs}{137} ratio in spent fuel could provide information on both the thermal
flux and the burn history.
The sensitivity to burn history is expressed in the ratio of the number of times the reactor was shutdown to the
the total irradiation time scaled by the mean lifetime of \nuc{I}{135}. This dimensionless factor encapsulates
the total dependence of the observed ratio on history.

\subsection{Effect of Reactor Shutdowns on the \nuc{Xe}{136}/\nuc{Xe}{134} Ratio}

The correction to the \nuc{Cs}{135}/\nuc{Cs}{137} ratio arising from reactor shutdowns, given by the second
term in Equation (\ref{cesium-ans}), implies an equivalent correction to the \nuc{Xe}{136}/\nuc{Xe}{134} ratio
of Equation (\ref{136-134}).


\begin{equation}
\frac{N_{136Xe}}{N_{134Xe}} = \frac{\overline{f}_{136Xe}}{\overline{f}_{134Xe}} \left[
    1 + \frac{\phi_T\sigma_a}{\lambda_{135Xe}+\phi_T\sigma_a} \left( \frac{\overline{f}_{135Xe}}{\overline{f}_{136Xe}} \right)
      \left(1 - \frac{P}{\lambda_{135I}T^{total}_{irrad}}
         \left(1 + \frac{\lambda_{135I}}{\lambda_{135Xe}+\phi_T\sigma_a}\right)\right)\right]\;.
\label{xe-anal}
\end{equation}

As mentioned above, the terms inversely proportional to
the total irradiation time arise from proper treatment of startup transients.
As can be seen from Equation (\ref{xe-anal}) and Figure  \ref{xenon-data}, the
reactor shutdown correction to the flux dependence for the \nuc{Xe}{136}/\nuc{Xe}{134} ratio
is small; this follows because of the way it appears in comparison to the constant 1 in the second term,
a qualitatively different form compared to the expression in Equation (\ref{cesium-ans}).

\section{Comparison of the Xenon and Cesium ratios to Experiment}

Maeck {\it et al.} measured \cite{Maeck} a series of fission isotope product ratios by irradiating highly
enriched uranium targets in the Advanced Test Reactor (ATR) and in the Engineering Test Reactor (ETR) at Idaho
National Laboratory.
Each target was exposed to a different reactor thermal flux by placing the targets at different distances from
the reactor mid-plane. The flux range was $6\times10^{12}-1.5\times10^{14}\; n/cm^2/sec$.
The targets loaded in the ETR were  more that 99\% enriched uranium-235. The targets were divided into two
groups, one that was irradiated for 20 days and one that was irradiated for 180 days over a 320 day period. The
targets in the  ATR were 93\% enriched and were irradiated for approximately 100 days over a period of 335 days,
with several reactor shutdowns over the 100 day irradiation period. The shape of the neutron flux for the
two reactors is different, with the capture to fission ratio for uranium-235 being $\alpha_{235}\sim 0.18$
for the ETR and $\alpha_{235}\sim 0.215$ for the ATR. The flux shape also varied very slightly from target
to target because of the difference in  location within the reactor. 

\begin{figure}
\vspace{0.8 cm}
\includegraphics[width = 8.5cm]{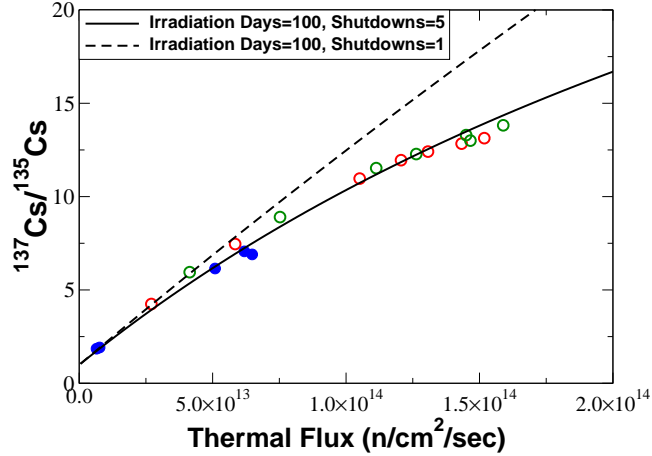}
\caption{The \nuc{Cs}{137}/\nuc{Cs}{135} ratio. 
The measurements are the data of Maeck {\it et al.}\cite{Maeck}. 
The blue circles are data from 99\% enriched uranium-235 samples that were irradiated in the ETR for either 20
days or 185 days (over a 320 day period). The green and red circles are data from 93\% enriched samples that
were irradiated in the ATR for 100 days over a period of 335 days. 
The dashed line is for a total irradiation time of 100 days, with one final shutdown. 
The straight line corresponds to an irradiation time of 100 days with a total of 5 reactor shutdowns. 
Both use the analytic expression (\ref{cesium-ans}).
We note that this latter scenario leads to a \nuc{Cs}{137}/\nuc{Cs}{135} ratio that is  identical to that for 20 days
irradiation, with one (final) shutdown, Equation (\ref{cesium-ans}). 
For high flux, it is clear that reactor shutdowns have a significant effect 
on the \nuc{Cs}{137}/\nuc{Cs}{135} isotopic ratio.} 
\label{cesium-data}
\end{figure}

Figure \ref{cesium-data} shows a comparison between the experimental data and the calculations for the
\nuc{Cs}{137}/\nuc{Cs}{135} isotopic ratio. (We note that, purely for display purposes, we follow Maeck et al.
\cite{Maeck} and always graph the cesium ratio as \nuc{Cs}{137}/\nuc{Cs}{135} and not the inverse.)
The  \nuc{Cs}{137}/\nuc{Cs}{135} ratio scales approximately linearly with the flux. Reactor shutdowns during the
irradiation can have a significant effect on the the cesium ratio at high flux, $\phi_T\gtrsim 5\times 10^{13}\;
n/cm^2/sec$.

\begin{figure}
\vspace{0.8 cm}
\includegraphics[width = 8.5cm]{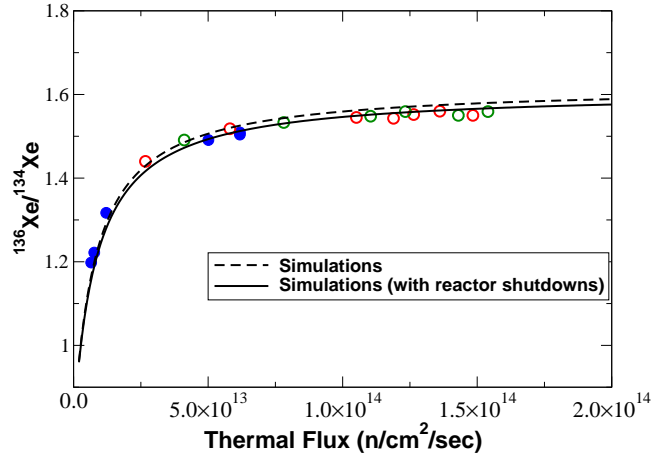}
\caption{The \nuc{Xe}{136}/\nuc{Xe}{134} ratio. 
The measurements are the data of Maeck {\it et al.}\cite{Maeck}. 
The dashed line is for a total irradiation time of 100 days, with one final shutdown. 
The straight line corresponds to an irradiation time of 100 days with a total of 5 reactor shutdowns. 
We note that this latter scenario leads to a \nuc{Xe}{136}/\nuc{Xe}{134} ratio that is  identical to that for 20 days
irradiation, with one (final) shutdown, Equation (\ref{cesium-ans}).  As can be seen, reactor shutdowns have
quite a small effect on the \nuc{Xe}{136}/\nuc{Xe}{134} isotopic ratio. This is in contrast to the effect of
shutdowns on the  \nuc{Cs}{137}/\nuc{Cs}{135} ratio.} 
\label{xenon-data}
\end{figure}

Figure \ref{xenon-data} shows the flux dependence of the \nuc{Xe}{136}/\nuc{Xe}{134} ratio. This ratio is also
a sensitive function of the flux, though less sensitive than the cesium ratio. The \nuc{Xe}{136}/\nuc{Xe}{134} varies
by a factor of $\sim2$ with the flux, Equation (\ref{limits}).  It  is most sensitive at low flux. Reactor
shutdowns only change the xenon ratio slightly.

\section{Reactor Burn Simulations}

\begin{figure}
\vspace{0.8 cm}
\includegraphics[width = 8.5cm]{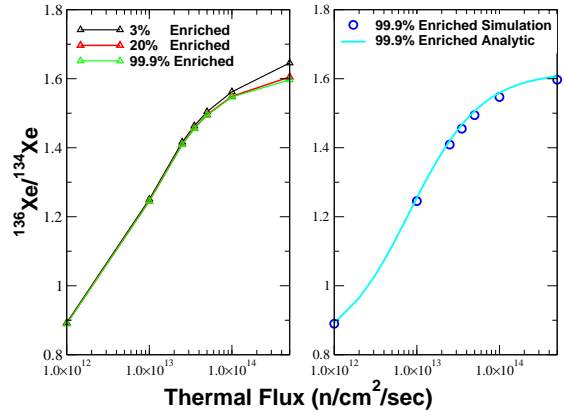}
\caption{Full reactor burn simulations for the \nuc{Xe}{136}/\nuc{Xe}{134} ratio.
The left panel compares the full simulations for three uranium-235 enrichments.
The right panel compares the results of the full reactor simulations with the analytic
expression of Equation (\ref{xe-anal}).
 As can be seen, the fuel
enrichment does not affect the relation  between the \nuc{Xe}{136}/\nuc{Xe}{134} ratio and the flux significantly, for
flux values below about $5\times 10^{13}n/cm^2/sec$. For the highest flux considered, $5\times 10^{14} n/cm^2/sec$,
the xenon ratio changed by 3\% in going from 3\% to 99.9\% enrichment. This effect is due to differences in the
plutonium growth rate and the corresponding changes in the cumulative fission yields ($\overline{f}_A$)
appearing in Equation (\ref{xe-anal}).}
\label{xe-enrich}
\end{figure}

\begin{figure}
\vspace{0.8 cm}
\includegraphics[width = 8.5cm]{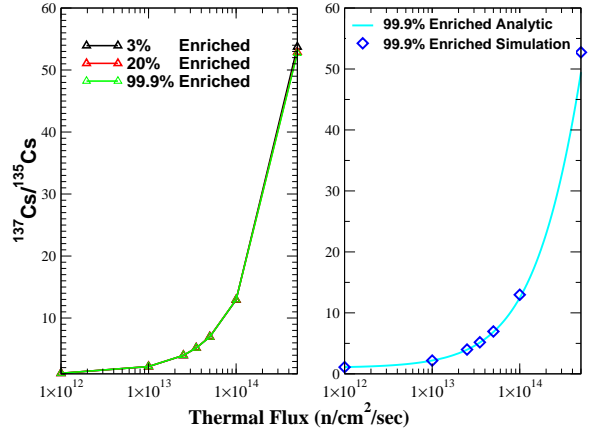}
\caption{Full reactor burn simulations for the \nuc{Cs}{137}/\nuc{Cs}{135} ratio.
The right panel compares the results of the full reactor simulations with the analytic expression of Equation
(\ref{cesium-ans}).
The left panel compares the full simulations for three uranium-235 enrichments. The fuel enrichment does not
affect the relation  between the \nuc{Xe}{136}/\nuc{Xe}{134} ratio and the flux significantly, except for the very
highest flux values.
At $\phi_T = 5\times 10^{14} n/cm^2/sec$, the cesium ratio changed by 2\% in going from 3\% to 99.9\%
enrichment.}
\label{cs-enrich}
\end{figure}

The previous sections of this paper use our analytic expressions for the xenon and cesium isotopic ratios. 
It is important, however, to validate these results using full reactor simulations. For this we ran a series of
simulations that included full production/depletion chains for all of the major actinides as well as the
tellurium, iodine, xenon, and cesium fission fragment chains. In these simulations we examined three uranium
fuel enrichments, 3\%, 20\%, and 99.9\% uranium-235 enrichment. In all cases the total irradiation time was
assumed to be 100 days, and we did not simulate shutdowns during the irradiation, since the shutdown
corrections are the same for the analytic and simulated results.
The results for the \nuc{Xe}{136}/\nuc{Xe}{134} isotope ratio are shown in Figure \ref{xe-enrich},
and those for the \nuc{Cs}{137}/\nuc{Cs}{135}  isotope ratio are shown in Figure \ref{cs-enrich}.
As can be seen, the analytic results provide a good representation of the full simulations. 
Both the xenon and cesium ratios are good indicators of the thermal flux, almost independent of the fuel
enrichment. The fuel enrichment becomes important at high flux where a significant fraction of the fissions
are from plutonium, for which the cumulative yields for xenon and cesium are  different from those for uranium.
But these effects change the xenon and cesium ratios by at most 3\%. 

\section{Inferring Information on Reactor Shutdowns}

\begin{figure}
\vspace{0.8 cm}
\includegraphics[width = 8.5cm]{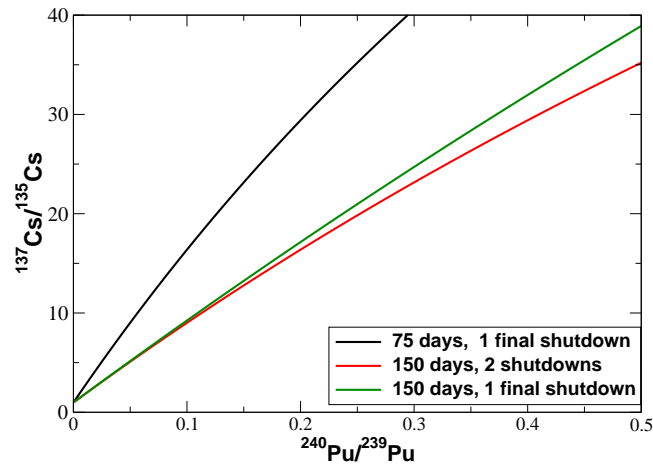}
\caption{The \nuc{Cs}{137}/\nuc{Cs}{135} ratio as a function of the \nuc{Pu}{240}/\nuc{Pu}{239} ratio clear breaks the
degeneracy between scenarios that involve the same ratio of number of reactor shutdowns to total irradiation
time ($P/\lambda_{135I}T_{irrad}^{total}$). Thus, a measurement of both ratios in spent reactor fuel would
make it possible to deduce the reactor flux and the number of times the reactor was shutoff over the time of
the fuel irradiation. We note however, that for low flux there is little sensitivity
to reactor shutdowns, and the cesium ratio directly determines the flux to which the fuel was exposed.}
\label{cs-pu}
\end{figure}

The correction to the cesium ratio from reactor shutdowns during the irradiation comes as a ratio of the total
number of shutdowns to the total irradiation time. Therefore there is a degeneracy of possible solutions in
relating the cesium ratio to the latter ratio.
For example,  a 20 day irradiation with one final shutdown would yield the same curve for flux versus cesium
ratio as an irradiation of 100 days with 5 shutdowns. To break this degeneracy requires the use of an additional
isotope ratio, and one straightforward possibility is to use a ratio that only depends on the thermal fluence or
the number of fissions. An ideal probe of the thermal fluence is the ratio of \nuc{Pu}{240}/\nuc{Pu}{239}, since this
ratio scales with the fluence but is not separately dependent on the flux or the enrichment of the in-going
uranium fuel \cite{poster}. In Figure \ref{cs-pu} we compare two scenarios, 75 days irradiation with one final
shutdown and 150 days with two shutdowns. It is clear that,  when plotted as a function of the plutonium ratio,
these two scenarios become clearly distinguishable. 
If undetonated fissile material were to be seized, for example, these two ratios could be used to infer
information on the reactor flux and the number of shutdowns. 

\section{Dilution of the Xenon Ratio into the Natural Atmosphere}
In situations were the xenon ratio is measured as an effluent from fuel reprocessing at a finite distance from the stacks of the
reprocessing facility,
the dilution and mixing of the released stable xenon isotopes into the natural background
of the air has to be taken into account.
In this section, we estimate the volume of air into which the fission gases can be diluted, while maintaining useful information on the reactor flux from the $^{136}$Xe/$^{134}$Xe ratio.

We consider the general case in which two gas samples $X$ and $Y$ are mixed to form a blend $B$, such that the blend contains $N_x$ and $N_y$ atoms from samples $X$ and $Y$, respectively, and $B=N_x/N_y$. 
If the ratio of any two stable xenon isotopes ${i,j}$ in the original samples  $X$ and $Y$
is $R_{ijX}$ and $R_{ijY}$, respectively,  
then the ratio in the blend $R_{ijB}$ is given by the dilution equation \cite{dilute1, dilute2}
\begin{equation}  
R_{ijB}=\frac{B\cdot R_{ijX}\sum_i R_{ijY} + R_{ijY} \sum_i R_{ijX}}{\sum_i R_{ijX} +B\cdot \sum_iR_{ijY}}
\end{equation}

The number of atoms of xenon gas released during reprocessing depends on both the mass of heavy metal reprocessed 
and on the total reactor irradiation time to which the reprocessed fuel was exposed; 
for a given neutron flux, the irradiation time determines
the grams of xenon per kilogram of heavy metal in the reprocessed fuel. The irradiation time can vary considerably,
depending on the application. For example, irradiation times for medical isotopes
are typically of the order of a few days, while those of weapons (reactor) grade plutonium
are months (years).
To estimate the effect of dilution on the xenon ratio, we ran reactor simulations for a burn history typical of what might be used for the production of weapons grade plutonium.  
We took the reactor flux to be $\phi = 10^{14}\; n/cm^2/sec$ and the $^{235}$U enrichment of the in-going uranium to be 3\%. 
The irradiation time was assumed to be $5\times 10^{6}\; secs$, 
corresponding to a total fluence of $5\times10^{20} n/cm^2$ and a $^{240}$Pu/$^{239}$Pu ratio of $\sim7\%$.

\begin{figure}
\includegraphics[width=12.5cm]{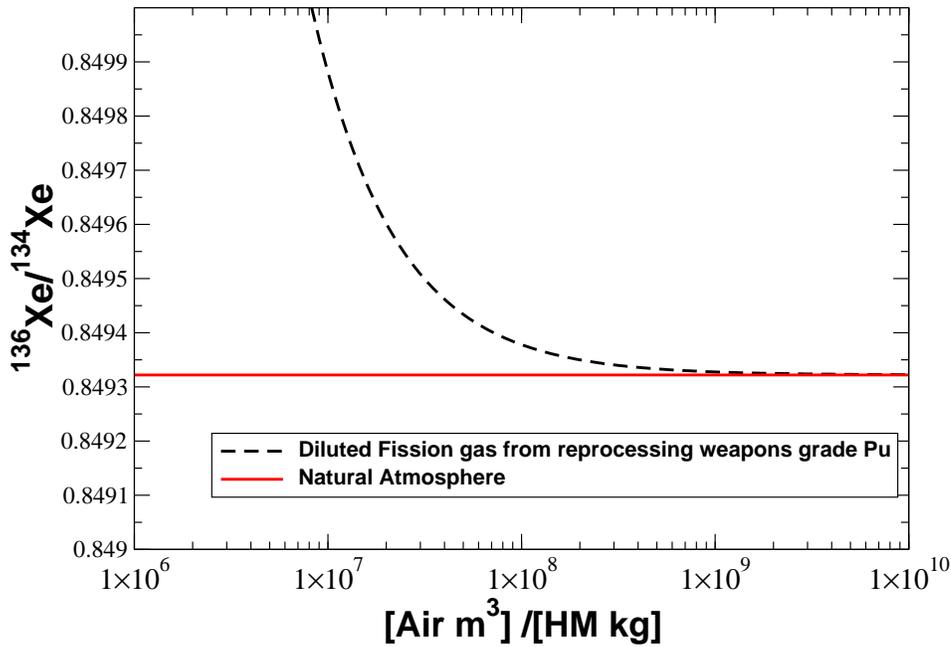}
\caption{The dilution of the fission produced $^{136}$Xe/$^{134}$Xe ratio from mixing with the natural xenon in the atmoshpere. 
The x-axis is the volume of air into which the fission gases are diluted per
kilogram of heavy metal (HM) being reprocessed.
Mass spectroscopic measurements of xenon ratios are accurate to better than 1 part in $10^4$. Thus, the $^{136}$Xe/$^{134}$Xe ratio retains reactor relavent information for dilutions into up to $10^9 m^3$ of air.  }
\label{dil1}
\end{figure}

Under these assumptions, the produced $^{136}$Xe/$^{136}$Xe ratio was 1.5799. The effect of dilution of the fission gases into the atmosphere on this ratio, as a function of the volume of air, is  shown in Figs. \ref{dil1}.
As can be seen, the xenon ratio retains reactor information for dilutions up to
about $10^9$ $m^3$ of air. 
A more detailed anaysis of the dilution problem for other reprocessing scenarios, as well as the important issue of determining the blend $B$, are the subject of a second paper \cite{our-dilution}.

\section{Conclusions}
The \nuc{Cs}{137}/\nuc{Cs}{135} and the \nuc{Xe}{136}/\nuc{Xe}{134} isotope ratios in spent reactor fuel provide direct
information on the  reactor flux to which the fuel was exposed. 
The cesium ratio shows the greater sensitivity to the flux and varies by a factor of 50 over the thermal flux
range $10^{12} - 5\times10^{14} n/cm^2/sec$. 
At the lowest flux values ($10^{12}\;n/cm^2/sec$), the cesium ratio is close to unity and approximately independent
of the flux. As the flux increases the relative concentration of cesium-135 is reduced by the transmutation of
xenon-135 to xenon-136 through rapid neutron capture.
If the age of the fuel is known, allowing corrections for the 30 year half-life  decay of cesium-137, the
\nuc{Cs}{137}/\nuc{Cs}{135} ratio is  always greater than unity. 
A decay-corrected observed ratio above unity is indicative of a reactor flux greater than $10^{12} n/cm^2/sec$. 
For large flux, the cesium ratio is significantly altered by reactor shutdowns during the irradiation of the
fuel. If the spent fuel can be analyzed in detail, the number of shutdowns can be inferred by comparing the
\nuc{Cs}{137}/\nuc{Cs}{135} ratio to the \nuc{Pu}{240}/\nuc{Pu}{239} ratio.
 
The \nuc{Xe}{136}/\nuc{Xe}{134} ratio varies by a factor of about 2 over the flux range $10^{12}- 5\times10^{14}
n/cm^2/sec$. If the flux and power of the reactor are varied over the burn history, the xenon ratio determines
the {\it average} flux to which the fuel was exposed. Thus, if xenon gases released in fuel reprocessing are
measured, the \nuc{Xe}{136}/\nuc{Xe}{134} ratio determines the physically relevant flux for reactor burn history
analyses. Finally, we note that the xenon ratio shows most sensitivity to the flux for the lower flux values,
whereas the cesium ratio is more sensitive in the higher flux range.

\section{Acknowledgement}
This work was funded through a research grant provided by the Department of Energy, National Nuclear Security
Agency, NA-22 office of Nonproliferation and Verification Research and Development.

\end{document}